# Virtual Machines Embedding for Cloud PON AWGR and Server Based Data Centres

Randa A.T. Alani, Taisir E.H. El-Gorashi, and Jaafar M.H. Elmirghani

*School of Electronic and Electrical Engineering, University of Leeds, LS2 9JT, United Kingdom*

**ABSTRACT**

Virtualization is one of the most active areas in cloud computing networks research. Passive Optical Networks (PONs) provide several benefits to the data centres such as low cost, low energy consumption and high bandwidth. In this study, we investigate the embedding of various cloud applications in PON AWGR and Server Based Data Centres. We optimize the power consumption of cloud applications' placement through the use of Mixed Integer Linear Programming (MILP). The results show that a reduction in power consumption by 24%, 22% and 26% can be achieved compared to non-optimized embedding of 5, 10 and 15 VMs respectively.

## 1. INTRODUCTION

Recent developments in cloud computing have increased the need for scalable, cost effective, high bandwidth, low power consumption data centre infrastructures [1] – [8]. The developments in data centre energy efficiency have accelerated making use of new techniques to improve the energy efficiency of core and access communication networks [9] – [17]. These energy efficient optical network architectures found their way to data centres. The increase in data intensive applications and processing requirements was an additional driver for growth in data centre demands [18] – [21]. Network virtualization technologies developed to offer better control of the network and better utilization of the network resources [22] – [24] became important candidates for the development of virtualized high data rate optical data centre networks. Several studies have proposed Passive Optical Networks (PONs) in data centre infrastructures due to their proven performance in access networks. With similar objectives, we introduced different novel PON architectures to handle the inter-rack and intra-rack communication of a data centre as in [25] – [33]. In [30] we studied one of the PON designs proposed in [29] where Arrayed Waveguide Grating Routers (AWGRs) and servers are used to route traffic and showed that power consumption is reduced by 83% compared to the Fat-Tree architecture [34] and by 93% compared to the BCube architecture [35]. In this paper, we further investigate this PON architecture for considering cloud applications. We developed a developed Mixed Integer Linear Programming (MILP) model that optimizes power consumption by optimizing the embedding of virtual machines' requests.

The reminder of this paper is structured as follows: In Section 2, the PON DCN Design with AWGR and Server Based Routing is briefly described. In Section 3, the power consumption results are presented. Finally, conclusions are provided in Section 4.

## 2. THE AWGR AND SERVER BASED-PON DCN ARCHITECTURE

In this architecture, as shown in Fig. 1, the servers in PON cells are placed into groups. A group is made of multiple subgroups where the number of servers hosted by a subgroup depends on the splitting ratio of the TDM PON connected to it. This architecture involves two types of communication: intra group and inter group communication. Intra group communication can be either between servers in a subgroup (intra subgroup communications) or between servers in different subgroups in the group (inter subgroup communication). The subgroups in each group are connected to a special server whose task is to maintain the inter subgroup and inter group communication

## 3. POWER OPTIMIZATION OF VIRTUAL MACHINES EMBEDDING IN PON DATA CENTER

A MILP model is developed to minimize the power consumption of a PON data centre shown in Fig.1 by optimizing the embedding of virtual machines requests. Two groups are considered where each group is divided into two subgroups. The number of servers in each subgroup is varied (2, 3,4) to study its effect on the total power consumption. Also, the two types of communication: intra group and inter group communication are covered by this reduced architecture. A number of VM requests (5, 10, and 15) were considered and Table 1 shows the different CPU, memory, and communication traffic demands of VMs that were studied. Each VM can have communication with 1-3 other VMs.

This model guarantees that the memory and processing demands allocated to a server do not exceed the server's capacity. Also, it ensures that the special server's processing capacity used for 'requests forwarding' does not exceed the special server's overall capacity.

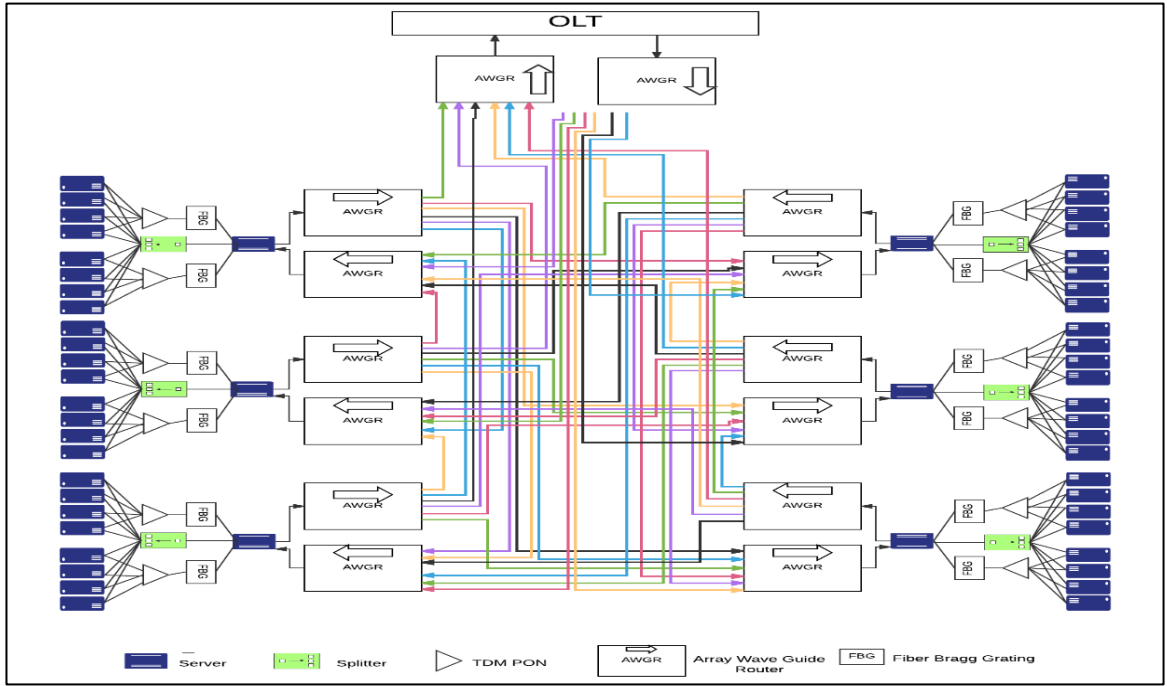

*Figure 1: Option 4 architecture*

Table1: Input data for the model

| Traffic demand between VMs | 40-200 Mb/s |
|---|---|
| Capacity of physical link | 10 Gbps |
| Idle power consumption of a server | 201 W [36] |
| Maximum power consumption of a server | 301 W [36] |
| Processing capacity requested by a client in *M* CPU cycles | 500-2000 |
| Processing capacity of server | 2.5 GHz |
| Portion of a special server's processing capacity used for forwarding one request | 5% |
| Processing capacity of special server | 2.5 GHz |
| Total ONU power consumption | 2.5 W [37] |
| ONU data rate | 10 Gbps |
| VM request requirements of RAM | 500-2000 MB |
| Memory capacity (RAM) of server | 8 GB |

In addition, traffic from servers in each PON group communicating with other servers is limited to the shared link capacity. Furthermore, we ensure that the traffic forwarded by a special server does not exceed its ONU rate. This model is used to minimize the total power consumption composed of servers, special servers and ONUs power consumption. This is achieved through optimizing the servers selected to provision VMs.

The power consumption of resource provisioning for the PON architecture is presented in Fig. 2. As shown in Fig. 2, the power consumption of placing VMs is inversely proportional to the number of servers in each subgroup (in general). The idea behind this is that whenever VMs that have traffic between them, are placed in one subgroup, there will be no need for activating the special servers connected to the subgroup. This more likely happens when the number of servers in a subgroup is high enough. Also, even if some VMs are distributed over more than one subgroup, the model tries to locate VMs with high traffic in the same subgroup as much as possible. Accordingly, a small amount of traffic traverses through the special servers connecting these VMs. This is important because the amount of power consumed by special servers is affected by the amount of traffic traversing them. In other words, whenever the number of servers is low, the VMs will distribute over more than one subgroup which activates the special servers and consumes more power. It worthy mentioning that increasing the number of servers in a subgroup is limited by the splitting ratio of the TDM PON connected to it.

It is clear from Fig. 3 that the number of activated servers for VMs placement is affected by the number of VMs allocated regardless of the number of servers' in each subgroup. In contrast, the number of activated 'special servers', changes according to change in the number of both VMs and servers in each group. This is shown in Fig. 4 where the change in the number of activated special servers is inversely proportional to the number of servers in each group and directly proportional to the number of placed VMs.

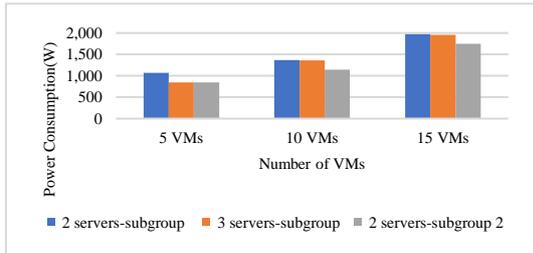

*Figure 2: VMs power consumption based on the number of servers in each subgroup*

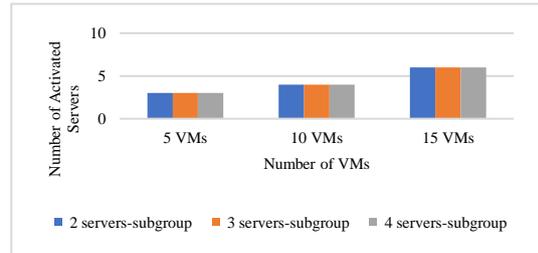

*Figure 3: Number of activated servers based on the number of servers in each subgroup*

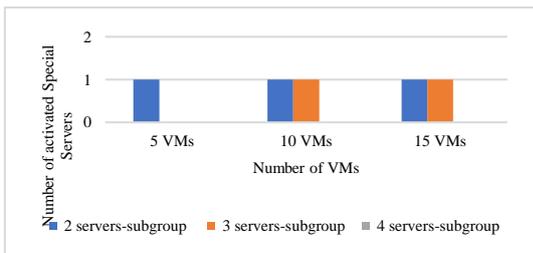

*Figure 4: Number of activated Special Servers based on the number of servers in each subgroup*

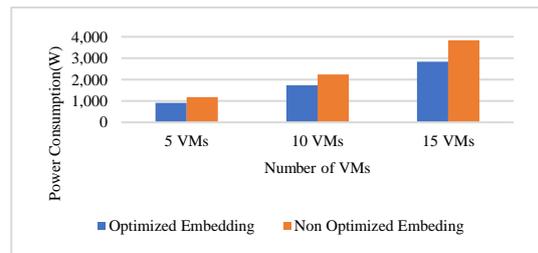

*Figure 5: VMs power consumption for optimized embedding and Non optimized embedding*

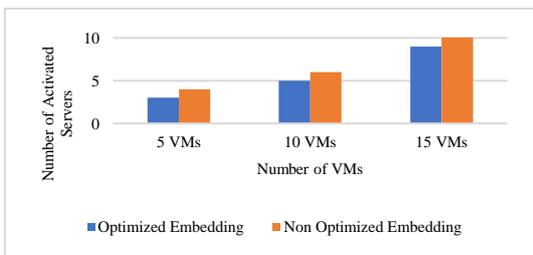

*Figure 6: Number of activated servers for optimzsed embedding and Non optimized VMs'embedding*

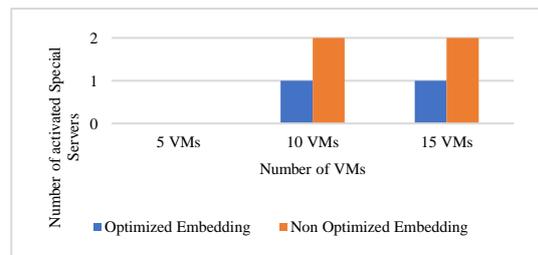

*Figure 7: Number of activated Special Servers for optimized embedding and Non optimized VMs'embedding*

Fig. 5 clearly shows that the optimized embedding model reduces the power consumption compared to the non-optimized embedding model. The main reason behind the energy savings is the minimized number of servers, 'special server' used for allocating VMs and providing communication as shown in Fig. 6 and Fig. 7. Furthermore, minimizing the traffic passing through the special servers reduces their operational power and ONU power consumption. This is done by allocating VMs that communicate with each other considering the following priority order: same server, subgroup or group as much as possible. Accordingly, our model succeeded in minimizing the power consumption by 24%, 22% and 26% compared to Non-optimized embedding model of 5, 10 and 15 VMs respectively.

## 4. CONCLUSIONS

This paper has investigated the placement of VMs in PON data centre architectures where Arrayed Waveguide Grating Routers (AWGRs) and servers are used to route traffic. We have optimized the power consumption of cloud applications allocation using a MILP model and presented a range of results. The results have shown that power consumption is affected by the number of servers in each subgroup due to the change in the number of activated special servers. Our study shows that the proposed model reduces the power consumption by 24%, 22% and 26% compared to Non-optimized embedding model of 5, 10 and 15 VMs respectively.

## ACKNOWLEDGEMENTS

The authors would like to acknowledge funding from the Engineering and Physical Sciences Research Council (EPSRC), INTERNET (EP/H040536/1) and STAR (EP/K016873/1). Mrs. Randa A.T. Alani would like to acknowledge The Iraqi Ministry of Higher Education and Scientific Research for funding her scholarship. All data are provided in full in the results section of this paper.